RESEARCH ARTICLE

# Skipping Selected Steps of DWT Computation in Lossless JPEG 2000 for Improved Bitrates

Roman Starosolski*

Institute of Informatics, Faculty of Automatic Control, Electronics and Computer Science, Silesian University of Technology, Gliwice, Poland

* rstarosolski@polsl.pl, rstaros@gmail.com

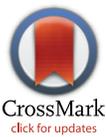

**Data Availability Statement:** Data is available on figshare at https://dx.doi.org/10.6084/m9.figshare.3457778.

**Funding:** This work was supported by the BK-219/RAU2/2016 grant from the Institute of Informatics, Silesian University of Technology and by the 02/020/RGH15/0060 grant from the Silesian University of Technology. The funders had no role in study design, data collection and analysis, decision to publish, or preparation of the manuscript.

**Competing Interests:** The authors have declared that no competing interests exist.

## Abstract

In order to improve bitrates of lossless JPEG 2000, we propose to modify the discrete wavelet transform (DWT) by skipping selected steps of its computation. We employ a heuristic to construct the skipped steps DWT (SS-DWT) in an image-adaptive way and define fixed SS-DWT variants. For a large and diverse set of images, we find that SS-DWT significantly improves bitrates of non-photographic images. From a practical standpoint, the most interesting results are obtained by applying entropy estimation of coding effects for selecting among the fixed SS-DWT variants. This way we get the compression scheme that, as opposed to the general SS-DWT case, is compliant with the JPEG 2000 part 2 standard. It provides average bitrate improvement of roughly 5% for the entire test-set, whereas the overall compression time becomes only 3% greater than that of the unmodified JPEG 2000. Bitrates of photographic and non-photographic images are improved by roughly 0.5% and 14%, respectively. At a significantly increased cost of exploiting a heuristic, selecting the steps to be skipped based on the actual bitrate instead of an estimated one, and by applying reversible denoising and lifting steps to SS-DWT, we have attained greater bitrate improvements of up to about 17.5% for non-photographic images.

## Introduction

In lossless JPEG 2000, the 5×3 kernel reversible discrete wavelet transform (DWT) performed using lifting steps decomposes an image into subbands of different characteristics, that are then independently entropy coded [1, 2]. In [3] we applied reversible denoising and lifting steps (RDLS) [4] to DWT; i.e., we integrated denoising into DWT lifting steps in such a way that the perfect reversibility of the transform was preserved despite the inherently lossy nature of denoising. We found that the noise filtering was the most effective in improving lossless JPEG 2000 bitrates when applied during computing of some RDLS-modified DWT (RDLS-DWT) subbands only. Since in some cases the best bitrates were obtained when the DWT stage of JPEG 2000 was skipped, we suspected that similarly to denoising, the optimum might be in-between skipping and applying DWT. Therefore, in this study we propose the skipped-steps DWT (SS-DWT), which is obtained from DWT by skipping selected steps of its computation. By selecting steps to be skipped we may obtain, as special cases of SS-DWT, an unmodified DWT (if we do not skip any step), skip the entire DWT (by skipping all the steps of the transform), or skip the DWT partially.

The bitrate improvements due to RDLS were attained in [3] at a cost, that might be too high for certain practical applications. E.g., an average improvement of bitrate of non-photographic images by almost 12% was obtained at the cost of a triple JPEG 2000 compression process execution with additional subband denoising. In this study we primarily focus on practical usefulness of the proposed methods. This approach results in finding a compression scheme that on average improves bitrates of non-photographic images by over 14% at the cost of compression time increased only by about 3%, as compared to the unmodified JPEG 2000. Another important practical property of this scheme is that it is compliant with the JPEG 2000 part 2 standard [5].

The remainder of this paper is organized as follows. In the next section, we briefly characterize the DWT in lossless JPEG 2000, the above-mentioned RDLS-DWT, the proposed SS-DWT along with the proposed basic heuristic for deciding which SS-DWT steps should be skipped, the application of RDLS to SS-DWT along with a heuristic for selecting denoising filters and deciding which steps to skip, and the experimental procedure. Then, subsections of section entitled "Investigations and discussion" present 4 stages of the research. First, based on SS-DWT effects for a large and diverse test image set, we choose the parameters of the basic heuristic (e.g., the number of iterations and the additional constraints for step-skip decisions). As the SS-DWT bitrate improvements are better, but generally similar to improvements obtained for the same data using RDLS-DWT, in the second stage we combine both methods and also compare them to JPEG-LS [6, 7] and HEVC [8, 9] in a lossless mode. Most of the bitrate improvement of combined SS-DWT and RDLS-DWT is obtained by using only the SS-DWT, which is simpler and more promising than RDLS-DWT with respect to both the compression ratio and the compression speed. As we focus on effects worthwhile from a practical standpoint, we further investigate SS-DWT only. In the third stage, based on the distribution of step-skip decisions of the basic heuristic, we propose a revised heuristic that has a lower computational time complexity and we define fixed SS-DWT variants that are compliant with the JPEG 2000 part 2 standard. In the final stage, to further reduce the cost of bitrate improvement, we test subband entropy as an estimator of JPEG 2000 encoding effects for the heuristic and for choosing among fixed variants. The last section summarizes the findings and indicates areas for future research.

## Materials and Methods

### Lifting-based DWT in lossless JPEG 2000

For brevity, as in the previous work [3], we describe here only the lifting-based 5×3 kernel reversible DWT that is exploited in baseline lossless JPEG 2000 compression of grayscale images, reduced to essentials. For further details as well as for more general characteristics of JPEG 2000, DWT, and the lifting scheme, the reader is referred to [1, 2, 5, 10–12].

Using the lifting scheme [11], the one-dimensional DWT (1D-DWT) transforms in-place a discrete signal $S = s_0 \, s_1 \, s_2 \ldots s_{l-1}$ of finite length $l$ into two subbands:

- a low-pass filtered signal $L$ that represents the original signal's low-frequency features;

- a high-pass filtered signal $H$ containing high-frequency features that, along with the low-pass signal, allows the perfect reconstruction of the original signal.

$S$ is transformed in 3 steps. First, in the prediction step, we perform the high-pass filtering of odd samples—hereafter, the parity of sample or pixel is determined by its location and not its value—by applying the lifting step (Eq 1) to each of them:

$$s_x \leftarrow s_x - \lfloor (s_{x-1} + s_{x+1})/2 \rfloor, \tag{1}$$

where the floor symbol $\lfloor v \rfloor$ denotes the greatest integer not exceeding $v$. Another lifting step is then applied to each even sample (update step):

$$s_x \leftarrow s_x + \lfloor (s_{x-1} + s_{x+1} + 2)/4 \rfloor. \tag{2}$$

Finally, in the reorder step, we reposition even samples to the lower half of the original signal, preserving their ordering (sample $s_x$ is moved to $s_{x/2}$), and odd samples are moved to the upper half. We obtain separate subbands $L$ and $H$, respectively. As opposed to prediction and update, the reorder step might be seen as an implementation detail of a JPEG 2000 coder, as it does not change the properties of the transformed samples within a given subband. However, the nearest neighbors of a sample are then used to determine its coding context in subsequent entropy coding; by changing positions of these samples we affect the context formation, which in turn affects the obtained bitrate.

The two-dimensional DWT (2D-DWT) for an image is obtained by first applying 1D-DWT to each image column, which results in $L$ and $H$ subbands of the image. Then by applying 1D-DWT to each row, we obtain the 1-level 2D-DWT, consisting of $LL$ and $HL$ subbands (transformed from $L$ subband) and $LH$ and $HH$ subbands (from $H$ subband); see Fig 1A–1C. We will call a subband belonging to a subband pair ($L$,$H$), ($LL$, $HL$), or ($LH$, $HH$) complementary to another subband from the same pair. The higher-level DWT, that provides multiresolution image representation, is obtained by Mallat decomposition [12]. The $t$+1-level transform is obtained by applying the $t$-level transform to the $LL$ subband of the $t$-level transform (Fig 1D). Mallat decomposition is the only one supported by the baseline JPEG 2000 (part 1) standard [1], noteworthy, the extensions of the baseline standard (Annex F of JPEG 2000 part 2) allow various arbitrary decomposition structures [5].

In lossless JPEG 2000, the transformed image is encoded in a complex and flexible manner [1, 2]. For the remainder of this study, it is noteworthy that each subband is compressed independently of the others using a context-adaptive entropy coder in which the context is subband-dependant.

## Application of Reversible Denoising and Lifting Steps to DWT

In [3], motivated by the observation that the lifting step of a color space transform may increase the amount of noise in an image component, we applied to the DWT the reversible denoising and lifting steps (RDLS) [4], which are lifting steps integrated with denoising filters. In brief, we replaced the prediction (Eq 1) and update (Eq 2) lifting steps with the RDLS-modified counterparts presented in below Eq 3 and Eq 4, respectively:

$$s_x \leftarrow s_x - \lfloor (s_{x-1}^d + s_{x+1}^d)/2 \rfloor, \tag{3}$$

$$s_x \leftarrow s_x + \lfloor (s_{x-1}^d + s_{x+1}^d + 2)/4 \rfloor, \tag{4}$$

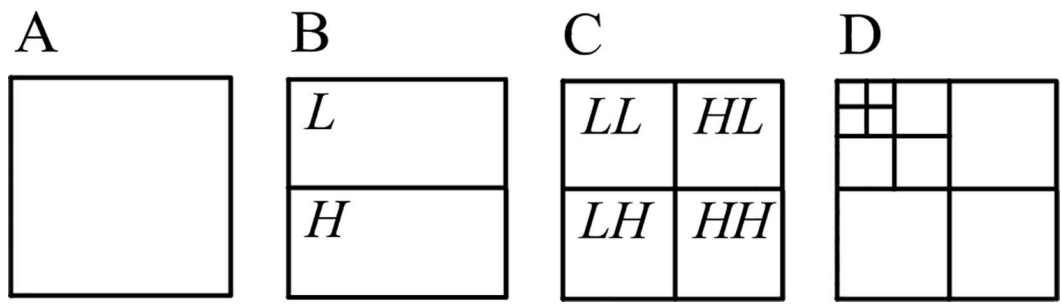

**Fig 1. 1-level 2D-DWT (A–C) and 3-level 2D-DWT (D).**

where $s_i^d$ is the denoised sample $s_i$ obtained using a deterministic denoising filter. Denoising is not an in-place operation, i.e., computing $s_i^d$ does not alter $s_i$. For the denoising of a sample of a specific parity, we used samples of the same parity only. The same denoising filter was used while computing all samples of a specific subband at a specific transform level. Despite the inherently lossy nature of denoising, RDLS exploiting denoising and transforms consisting of several such steps are perfectly invertible. For further details and RDLS examples the reader is referred to [3, 4, 13].

We found, that in the RDLS-modified DWT (RDLS-DWT) the noise filtering significantly improved the lossless JPEG 2000 bitrates of non-photographic images and of images contaminated with impulse noise; filtering was the most effective when applied to some subbands only. Since in some cases the best bitrates were obtained when the DWT stage of JPEG 2000 was skipped, we suspected that similarly to denoising, the optimum might be in-between skipping and applying the DWT, which in turn led to proposing the DWT modification presented in the following subsection.

## Proposed DWT modification

In this study, we evaluate the effects of skipping of selected steps of DWT computation on the lossless JPEG 2000 bitrates. In order to permit skipping the entire DWT we should allow to skip all the steps: the lifting-based prediction and update, and the reordering of samples. A transform in-between skipping and applying DWT is obtained by skipping some of the prediction, update, and reorder steps. However, as explained below, not all the possible combinations of performing or skipping various kinds of steps are practically justified. We also assume, that the subband characteristics are invariant and apply the same decision, as to perform or skip a specific step, to all samples of a given subband. In particular, for each level of DWT up to 9 binary decisions (to skip or to perform the steps) may be made. During computation of each of $H$, $LH$, and $HH$ subbands, we may either skip or perform the prediction lifting steps; for each of $L$, $LL$, and $HL$, we may skip or perform the update steps; and for each pair of complementary subbands ($L,H$), ($LL$, $HL$), and ($LH$, $HH$), if both prediction and update are skipped, we may skip or perform the sample reordering. We do not consider skipping the sample reordering if prediction or update was performed, because it would force the JPEG 2000 entropy coder to model and encode two sets of samples having different characteristics as if it was one homogeneous set and result in the bitrate worsening. Thus, at each transform level, for all samples of each pair of complementary subbands, we may:

- perform prediction, update, and reorder, or

- skip prediction and perform update and reorder, or

- perform prediction, skip update, and perform reorder, or

- skip prediction and update, and perform reorder, or

- skip prediction, update, and reorder.

Skipping the transform steps may make the characteristics of the transformed image subbands different to what is generally expected after DWT. We may obtain a subband containing a scaled down, but unfiltered, original image; if the reorder step gets skipped, then certain subbands may be not created (e.g., $LH$ and $HH$), making the baseline JPEG 2000 encoding less efficient due to entropy coding of such subband as of 2 separate non-existing subbands. On the other hand, in practice we are most interested in the bitrate that sometimes is the best when DWT is skipped. We denote the modified DWT with skipped steps as SS-DWT.

SS-DWT may be seen as a combination of two modifications of DWT that are known to be beneficial for bitrates: non-Mallat DWT decomposition structures (as skipping prediction, update, and reorder results in a non-Mallat decomposition) and using arbitrary wavelet kernels (as replacing the prediction or update lifting step with $s_x \leftarrow s_x$ is an effective equivalent of skipping the step). Both modifications are supported by the JPEG 2000 part 2 standard, however, in a certain extent only. E.g., we may define an arbitrary wavelet kernel for the entire DWT, and skip this way all predictions or all updates. Therefore, special SS-DWT cases may be compatible with this standard. On the other hand, SS-DWT is in some respects similar to nonlinear wavelet transforms. The nonlinearity of such a transform results from employing the wavelet kernel prediction filter (which determines the prediction lifting step) that is adaptively selected from a set of linear filters; e.g., see [14, 15]. In SS-DWT, we select the prediction step to be either Eq 1 or $s_x \leftarrow s_x$ and apply it to all samples in a subband. However, these nonlinear approaches neither employ lifting steps that effectively result in skipping of prediction or update nor allow to skip the reorder step.

Since testing the compression effects of all the possible combinations of binary decisions would be too complex even for just a few transform levels, we employed a step-skip selection heuristic. The basic heuristic is described in the following subsection and its revised variant in the next section. The resulting decisions should be transmitted to the decoder as a side information along with the compressed image. The size of this data is negligible. We need to transmit up to 9 binary decisions for each SS-DWT level, e.g., up to 27 bits for a 3-level SS-DWT.

## Basic step-skip selection heuristic

The basic heuristic is based on one used in [3] for the selection of denoising filters. It consists of the greedy steps described below, in which step B may be repeated for a certain number of iterations. For brevity, we describe a variant in which we assume that the reorder step is skipped if and only if both the prediction and update steps are skipped. Therefore, there is only one binary decision to be made for each subband regarding either the prediction or the update step only. By inverting such a decision we mean changing the decision either from performing to skipping the step or from skipping to performing it. The steps of the heuristic are as follows:

1. Perform JPEG 2000 compression of an image using DWT and using SS-DWT with skipping all of the steps. Then, to all subbands at all transform levels, assign the decision that resulted in a better bitrate.

2. For each transform level (starting from level 1) and for each subband (at the specific level analyzed in the $H$, $L$, $HL$, $HH$, $LL$, and $LH$ order), check if the overall bitrate is improved by inverting the decision for this subband and level; if it is improved, then invert the decision.

We denote this variant of basic heuristic as BH; when the number $n$ of iterations of step B is known, we follow the heuristic abbreviation with $(n)$, e.g., BH(1). In this study we also test the following variants of basic heuristic: the heuristic with additional testing if doing the reorder step improves bitrate when the prediction and the update is skipped (denoted BH_TR), the heuristic with disabled skipping of the reorder step (BH_AR), and the heuristic that for each pair of complementary subbands either skips all 3 DWT steps or performs them all (BH_PW). With a slight abuse of terminology, we also use the term JPEG 2000 compression for JPEG 2000 with DWT replaced by SS-DWT, which in a general case is not compliant with the JPEG 2000 standard, and without further modifications may disable certain JPEG 2000 functionalities, such as random code stream access for partial image decoding. Special SS-DWT cases that are compliant with the JPEG 2000 part 2 standard are discussed in the next section.

Subbands are encoded independently. In the heuristic step B we check the overall SS-DWT transformed image bitrate after changing the decision for a specific subband. The decision will only affect a subset of the subbands, therefore only some subbands need to be computed and encoded. For example, changing the decision for a level 2 subband does not affect the level 1 subbands. To get the overall bitrate in such case, we neither need to compute again any of the level 1 subbands ($H$, $L$, $HL$, $HH$, $LL$, or $LH$) nor to encode again the level 1 subbands $LH$, $HL$, and $HH$. There are two main elements of BH computational time complexity: performing the lifting steps (Eq 1 and Eq 2) and the encoding of transformed samples. Calculating them we took into account the above-mentioned property.

The number of symbols that need to be encoded by a single iteration of the heuristic step B for the 1-level transform and $p$-pixel image is $4p$, for the $t$-level transform it is $p \sum_{i=1}^{t} 4\left(\frac{1}{4}\right)^{i-1} = \frac{16}{3}(1 - 4^{-t})p$ and thus for BH($n$) it is

$$\left(2 + \frac{16}{3}(1 - 4^{-t})n\right)p,$$ (5)

whereas, obviously, the unmodified JPEG 2000 encodes $p$ symbols while compressing such image.

The 1-level DWT is done using $2p$ lifting steps and the $t$-level DWT requires

$$\frac{8}{3}(1 - 4^{-t})p$$ (6)

lifting steps. In the worst case, BH(1) for the level 1 of the $t$-level SS-DWT performs less than $6p+4a$ lifting steps, where $a$ is the number of lifting steps performed by DWT for transform levels from 2 to $t$; the overall number of lifting steps performed by BH($n$) is less than

$$\left(\frac{8}{3}(1 - 4^{-t}) + \left(\frac{104}{9} - \frac{1}{3}2^{5-2t}t - \frac{13}{9}2^{3-2t}\right)n\right)p.$$ (7)

Eqs 5–7 for increasing $t$ quickly converge to maxima. For infinite $t$, they are the upper complexity bounds of the heuristic (Eq 5 and Eq 7) and of DWT (Eq 6). Using these bounds, we may conveniently express the BH complexity as relative to the operations of the unmodified JPEG 2000. For infinite $t$, by dividing Eq 5 by $p$ and dividing Eq 7 by Eq 6, we find that the BH ($n$) computational time complexity is smaller than

$$\left(2 + \frac{16}{3}n\right)T_E + \left(1 + \frac{13}{3}n\right)T_D,$$ (8)

where $T_E$ is the complexity of encoding of the transformed image by JPEG 2000 entropy coder, and $T_D$ is the complexity of DWT computation. Note that we ignore the cost of the reorder step of DWT, which for 1-level 1D-DWT requires a single read and single write per sample but may be substituted with a transform-aware addressing of samples, whereas the lifting step requires 3 reads, a single write, and 3 (Eq 1) or 4 (Eq 2) arithmetic operations. Note also that for the final JPEG 2000 encoding of SS-DWT of an image based on decisions selected by the heuristic, the transformed image and encoded subbands are already generated by the heuristic and do not need to be computed again.

## Combining RDLS with step skipping

In [1], for an image-adaptive selection of denoising filters to be applied during computation of RDLS-DWT of a specific image, we used the heuristic consisting of the steps A and B presented below, in which step B may be repeated given a number of iterations.

1. For each of the denoising filters, perform JPEG 2000 compression of an image, using this filter in RDLS steps for all subbands at all levels. Then for all subbands at all transform levels, select the filter that resulted in the best overall bitrate.

2. For each transform level $a$ (starting from level 1) and for each subband $b$ (at a specific level analyzed in the $H$, $L$, $HL$, $HH$, $LL$, and $LH$ order), try to find a better denoising filter by checking for each filter (except for the one already selected) the bitrate obtained using this filter for subband $b$ at level $a$, while the filters selected so far are used for other subbands.

Experiments were performed for various denoising filters. We employed the special filter case, named the None filter, for which $s_i^d = s_i$; this filter turns RDLS into a regular lifting step (compare Eqs 1 and 2 with Eqs 3 and 4). Out of other regular denoising filters tested, linear and nonlinear, the latter were found to be effective in improving the bitrate of lossless JPEG 2000 and the Median denoising filter with 5×5 pixel window was the best.

To combine RDLS-DWT and SS-DWT, we employ another special RDLS filter case, named Null, for which $s_i^d = 0$; it was proposed in [13]. The Null filter results in skipping of the RDLS-modified prediction (Eq 3) or update (Eq 4) step, as both these steps become $s_x \leftarrow -s_x$. Additionally, if the Null filter is selected by the heuristic for both complementary subbands, i.e., if prediction and update are skipped, then we skip the reorder step as well. We denote this heuristic as H_SS_RDLS. The heuristic used with None and Null filters results in exactly the same transform, as obtained with BH. Obviously, by using it without the Null filter we get RDLS-DWT exactly as in [3]. By using None, Null, and the regular denoising filters we combine SS-DWT with RDLS-DWT; the heuristic for each lifting step may decide to keep it unchanged, skip it, or apply to it RDLS with an actual denoising filter. The resulting transform is denoted as RDLS-SS-DWT.

## Experimental procedure

In experiments, we used the green components of images from a "CT2" set [16]. The CT2 is a recent, large set of color images that was used in the research on lifting-based color space transforms [17, 18] and in our previous research [3]. It contains 746 images taken from different sources, and image sizes vary from 180x117 to 6600x5100. The set was divided into subsets: Photo, consisting of 499 photographs, and No-photo, consisting of 247 non-photographic computer-generated, screen content, or mixed-content images. The latter were further divided into images that were better compressed by JPEG 2000 without DWT (No-photo (a) contains 81 images) and with the unmodified DWT (No-photo (b) contains 166 images). The Photo images were not divided this way since only one such image was better compressed by JPEG 2000 without DWT than with the unmodified DWT.

We used the IRIS-JP3D JPEG 2000 part 10 (JP3D) [10, 19] reference software developed by Tim Bruylants from Vrije Universiteit Brussel (VUB) and the Interdisciplinary Institute for BroadBand Technology (IBBT), version 1.1.1 [20], which is downward compatible with the baseline JPEG 2000 standard. In this implementation it was relatively easy to modify DWT. Our implementation of SS-DWT (and of RDLS-DWT and RDLS-SS-DWT) is available [21] as a patch to the IRIS-JP3D. The codec executable was a single-threaded application compiled using the GCC-MinGW32 compiler, version 4.8.1. Experiments were performed on a computer equipped with Intel Xeon E3-1240 v.3 (3.40 GHz) processor and 16 GB RAM. In the experiments, except for replacing DWT with SS-DWT and setting the transform level, we used the default codec settings. The whole image was compressed as a single tile, and we used the 3-level decomposition. To switch off the DWT stage in the JPEG 2000 coding, we invoked the codec with the 0-level DWT setting.

To put the results of the JPEG 2000 variants in a wider perspective, we include in the comparisons the JPEG-LS [6, 7] and HEVC(H.265) [8, 9] compression algorithms in a lossless mode. The former is a standard of the ISO/IEC and ITU-T for primarily lossless compression of still images. The latter is the most recent video compression standard of ISO/IEC and ITU-T, which allows lossless compression of individual still images. JPEG-LS and HEVC results reported in the "Comparison with RDLS and other techniques" section were extracted from [3] and were obtained using the SPMG/UBC JPEG-LS implementation, version 2.2 [22] and the HEVC Test Model (HM) reference software, version 16.6 [23].

The compression ratio or bitrate $r$, expressed in bits per pixel (bpp), is calculated using the total size in Bytes of the compressed image that includes the compressed file format header. We ignored the cost of signaling to the decoder the lifting steps to be skipped, as on the average this cost was below 0.0001 bpp. We introduced modifications to JPEG 2000, and then we evaluated their effects by analyzing the obtained bitrate changes with respect to the bitrate of the reference method, that is, of unmodified JPEG 2000. The bitrate change $\Delta r$ was expressed in percentage of the reference method bitrate. The $\Delta r$ was also employed for comparisons with other methods. Due to the large number of images in our test-set, in Tables we report averaged bitrates and averaged bitrate changes for a set and for its specific subsets rather than results for individual images, whereas some figures present individual images' results.

The memoryless entropy $H_0$ of the $t$-level DWT or SS-DWT transformed image was calculated as a sum of memoryless entropies of all subbands that would be independently encoded by baseline JPEG 2000 with an unmodified $t$-level DWT, i.e., 10 subbands for 3-level transform, regardless of skipped reorder steps. Subband entropy, calculated as $-\sum_{i=0}^{N-1} p_i \log_2 p_i$, where $N$ is the alphabet size, and $p_i$ is the probability of occurrence of a sample value $i$ in the subband, was weighted with the size of the subband. $H_0$ was employed as an estimator of the JPEG 2000 subband encoding effects and used for fast entropy-based selecting of SS-DWT steps to be skipped by the heuristic and selecting among fixed SS-DWT variants. Thus $H_0$ was used only for constructing SS-DWT and only by some variants of the proposed method; the reported bitrates ($r$) and bitrate changes ($\Delta r$) are in each case calculated from the actual compressed image file sizes.

## Investigations and Discussion

### Observations for the basic heuristic

In Table 1, we present average changes of JPEG 2000 bitrate due to SS-DWT. For the BH, we see that after a single iteration of the heuristic step B, we obtain an average bitrate improvement of 5.55%, which is mainly due to the improvement for No-photo images, since bitrates of Photo images on average are improved by 0.75%. Using 2 iterations of step B as compared to the 1 iteration resulted in small, but noticeable further bitrate improvement of 0.21 percentage points for the whole test-set; increasing the number of iterations above 2 did not improve bitrates noticeably.

The obtained bitrate improvements are better than we expected. In the previous work [3], by using a more complex RDLS-DWT that involved testing several denoising filters, and by allowing to skip the DWT stage of JPEG 2000, for the same images we obtained average improvements of up to 5.21% and 0.65% respectively. Is skipping a DWT lifting step a more

**Table 1. Effects of SS-DWT on lossless JPEG 2000 bitrates for step-skip decisions selected by the BH and a couple of its variants.**

| Heuristic variant | Images | | | | |
|---|---|---|---|---|---|
| | **Photo** | **No-photo** | **No-photo(a)** | **No-photo(b)** | **All** |
| $r_{DWT}$ | 3.9975 | 2.9162 | 3.2882 | 2.7348 | 3.6395 |
| $\Delta r_{BH(0)}$ | -0.01% | -7.08% | -21.59% | 0.00% | -2.35% |
| $\Delta r_{BH(1)}$ | -0.75% | -15.25% | -32.26% | -6.96% | -5.55% |
| $\Delta r_{BH(2)}$ | -0.86% | -15.66% | -32.28% | -7.54% | -5.76% |
| $\Delta r_{BH(3)}$ | -0.86% | -15.66% | -32.28% | -7.55% | -5.76% |
| $\Delta r_{BH(4)}$ | -0.86% | -15.66% | -32.28% | -7.55% | -5.76% |
| $\Delta r_{BH\_TR(1)}$ | -0.75% | -15.28% | -32.30% | -6.98% | -5.56% |
| $\Delta r_{BH\_TR(2)}$ | -0.86% | -15.69% | -32.33% | -7.57% | -5.77% |
| $\Delta r_{BH\_AR(1)}$ | -0.73% | -13.24% | -27.26% | -6.40% | -4.87% |
| $\Delta r_{BH\_AR(2)}$ | -0.74% | -13.38% | -27.59% | -6.44% | -4.93% |
| $\Delta r_{BH\_PW(1)}$ | -0.31% | -9.26% | -23.35% | -2.38% | -3.27% |
| $\Delta r_{BH\_PW(2)}$ | -0.31% | -9.29% | -23.43% | -2.40% | -3.28% |

$r_{DWT}$–average lossless JPEG 2000 bitrate for 3-level unmodified DWT (bpp); $\Delta r_{variant}$–average SS-DWT bitrate change obtained with use of the heuristic variant: BH–basic heuristic, BH_TR–heuristic with additional testing if doing the reorder step improves bitrate, BH_AR–heuristic with disabled skipping of the reorder step, BH_PW–heuristic that for each pair of complementary subbands either skips all 3 DWT steps or performs them all.

effective and a simpler method of bitrate improvement, that should be used instead of replacing the step with RDLS? Or otherwise, maybe for different steps either method is the best and by combining SS-DWT and RDLS-DWT we will obtain further bitrate improvements large enough to justify the increased complexity of such a method? We investigate this in the next subsection.

We also tested a couple of variants of the BH and found them to not be worthwhile (Table 1). When both prediction and update steps are skipped, by additional checking if performing the reorder step improves the bitrate we get a negligible further bitrate improvement at the cost of increased heuristic complexity. Unconditionally performing the reorder step results in smaller improvements of below 5%; the impact of the reorder step on the obtained bitrate is not negligible. Simplifying the heuristic by either performing or skipping all 3 DWT steps considered together for each pair of complementary subbands decreases the average improvement to over 3%.

We expected that the minimum bitrate may be obtained when only some DWT steps are skipped. Indeed, for the majority of images (702 out of 746), the decisions obtained by BH(2) are to skip some steps only; only for 38 images the decisions are to not skip any steps, and only for 6 No-photo images–to skip all the steps. For the latter, by skipping the DWT stage of JPEG 2000 instead of using the 3-level SS-DWT with all the steps skipped, which results in JPEG 2000 treating the image as a single subband, we would decrease the bitrate by 1.01%. The above cost is not significant when we consider the entire test-set. For another 6 images, although skipping some steps results in the best bitrate when the transformed image is encoded as if it was transformed by 3-level DWT, a lower bitrate is obtained when the DWT stage is skipped. Overall, the average set bitrate may be decreased by exploiting the DWT stage skipping by 0.02%.

## Comparison with RDLS and other techniques

In this section we aim to identify the JPEG 2000 variant employing step skipping, RDLS, or both methods that is the most worthwhile from a practical standpoint. In Table 2, we compare average SS-DWT improvements of lossless JPEG 2000 bitrate to improvements obtained using

**Table 2. Results for SS-DWT, RDLS-DWT, RDLS-SS-DWT, and comparison with JPEG-LS and HEVC (lossless mode).**

| Compressor variant | Images | | | | |
|---|---|---|---|---|---|
| | **Photo** | **No-photo** | **No-photo(a)** | **No-photo(b)** | **All** |
| $r_{DWT}$ | 3.9975 | 2.9162 | 3.2882 | 2.7348 | 3.6395 |
| $\Delta r_{BH(2)}$ | -0.86% | -15.66% | -32.28% | -7.54% | -5.76% |
| $\Delta r_{RDLS(2, 2)}$ | -0.57% | -12.68% | -27.18% | -5.61% | -4.58% |
| $\Delta r_{RDLS(5, 2)}$ | -0.64% | -13.29% | -28.45% | -5.89% | -4.83% |
| $\Delta r_{RDLS(2, 2), NO-DWT}$ | -0.58% | -14.06% | -31.37% | -5.61% | -5.04% |
| $\Delta r_{RDLS(5, 2), NO-DWT}$ | -0.65% | -14.43% | -31.92% | -5.89% | -5.21% |
| $\Delta r_{RDLS-SS(3, 2)}$ | -0.86% | -17.20% | -36.36% | -7.85% | -6.27% |
| $\Delta r_{RDLS-SS(6, 2)}$ | -0.92% | -17.47% | -36.82% | -8.03% | -6.40% |
| $\Delta r_{JPEG-LS}$ | -2.72% | -21.46% | -38.27% | -13.26% | -8.92% |
| $\Delta r_{HEVC}$ | 9.32% | -17.04% | -36.48% | -7.55% | 0.59% |

$r_{DWT}$—average lossless JPEG 2000 bitrate for 3-level unmodified DWT (bpp); $\Delta r_{variant\_list}$—average bitrate change obtained using the best out of the listed variants: BH($n$)—SS-DWT with step-skip decisions obtained using BH, RDLS($f$, $n$)—RDLS-DWT with denoising filters selected by the H_SS_RDLS heuristic, RDLS_SS($f$, $n$)—RDLS-SS-DWT using H_SS_RDLS, $n$—number of iterations of step B of the heuristic, $f$—number of denoising filters, NO-DWT—skipping the DWT stage of JPEG 2000; $\Delta r_{JPEG-LS}$—average bitrate change obtained by using JPEG-LS instead of JPEG 2000; $\Delta r_{HEVC}$—average bitrate change obtained by using HEVC.

doi:10.1371/journal.pone.0168704.t002

RDLS-DWT and RDLS-SS-DWT. Step-skip decisions and denoising filters were selected for each image individually in 2 iterations of step B of the respective heuristic. For RDLS-DWT, we additionally allowed to skip the entire DWT stage of JPEG 2000. For transforms involving denoising, we tested both the minimum-size sets of denoising filters and greater sets. The minimum-size set for RDLS-DWT contained 2 filters: None and Median with 5×5 pixel window; in the case of RDLS-SS-DWT it contained also the Null filter. Larger sets contained the None filter, 4 nonlinear denoising filters (including above-mentioned Median and 3 milder filters described in [3]), and in the case of RDLS-SS-DWT the Null filter.

As already noted, skipping of selected DWT steps results in bitrate improvements greater than obtained by applying RDLS to DWT—even when for the latter method we allow to skip the DWT stage of JPEG 2000 and use several denoising filters. By combining both methods, however, we obtain bitrate improvements, that are better than in the case of each of them considered alone. Using RDLS-SS-DWT with just one actual denoising filter (see Table 2 row: $\Delta r_{RDLS-SS(3, 2)}$) we obtain average improvement for all the images greater by 0.5 percentage points than obtained using SS-DWT. By employing more denoising filters we get a small further improvement. Noteworthy, RDLS-SS-DWT with minimal number of filters results in much better bitrates, than RDLS-DWT with a greater filter set.

In Table 2, we also compare the effects of step skipping and RDLS on lossless JPEG 2000 bitrates to two other compression algorithms in a lossless mode: JPEG-LS and HEVC(H.265). Let's look at the results of JPEG-LS and HEVC, starting from the No-photo images. Both JPEG-LS and HEVC are for these images significantly better than the unmodified JPEG 2000. By applying SS-DWT we obtain bitrates, that are much closer to JPEG-LS and HEVC, than to unmodified JPEG 2000. Using RDLS-SS-DWT gives a further improvement, that is much smaller, than obtained by an application of SS-DWT to JPEG 2000. However, this smaller step allows to attain bitrates better than in the case of HEVC algorithm. For Photo images, HEVC is worse than unmodified JPEG 2000, whereas JPEG-LS is better by 2.72%, i.e. by much less than for No-photo images (21.46%). By modifying JPEG 2000 we obtain only a small fraction of improvement possible for these images by use of JPEG-LS. On the other hand, JPEG-LS results indicate that a further improvement of lossless JPEG 2000 bitrates should be possible.

SS-DWT results in majority of the bitrate improvement attainable with significantly more complex RDLS-SS-DWT; as compared to RDLS-DWT it is better with respect to both the bitrate improvement and the complexity. It allows for significant improvement of the JPEG 2000 bitrate as compared to JPEG-LS and HEVC. All in all, it is the most promising variant from a practical standpoint; thus in next subsections we focus on improving it further. We also note that greater bitrate improvements are attainable by combining RDLS-DWT and SS-DWT and leave investigating RDLS-SS-DWT as an interesting topic for future research.

## A revised heuristic and fixed SS-DWT variants

Interesting observations may be made based on analyzing how often the prediction or update and reorder steps are skipped for specific subbands and SS-DWT levels (Fig 2). The distribution of decisions is similar for the same subband at different transform levels. However, at level 1 of transform for Photo images, it is usually better to skip the update step (subbands $L$, $LL$, and $LH$), whereas for higher levels more or equally frequently, it is better to perform the update. Nearly always (except for 2 images out of 746), if prediction is skipped, then the update for the complementary subband and reorder are also skipped. Thus, we may simplify the heuristic—for a given level and a pair of complementary subbands, if the prediction is skipped, then we should also skip the update step.

Our test-set is divided into Photo and No-photo images based on *a priori* knowledge, which is not available to the heuristic. However, based on the heuristic step A outcome, we may assume that the image is No-photo (a), i.e., that it is better compressed with the switched off DWT stage in the JPEG 2000, or other (we would be wrong for 2 images only). In Fig 2, for No-photo (a) images, we can see that the update steps (potentially performed for $L$, $LL$, and

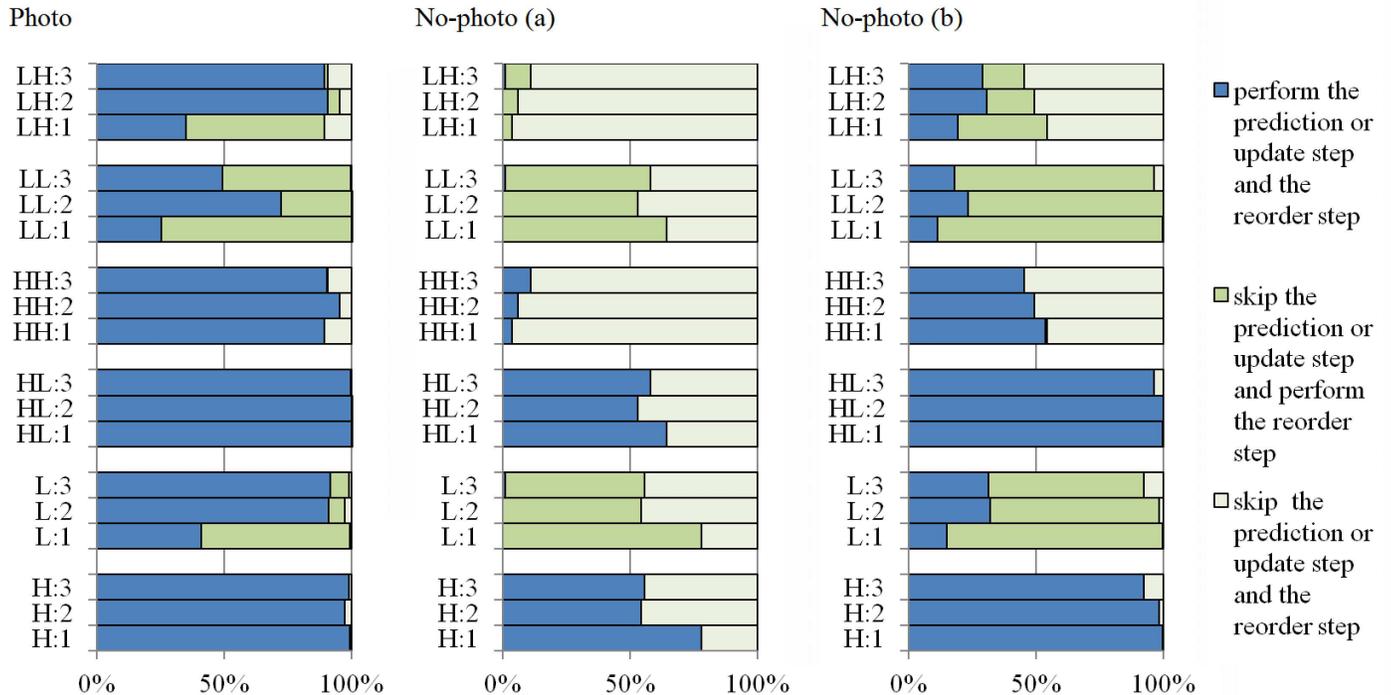

**Fig 2. SS-DWT step-skip decisions selected by BH(2) for Photo (left-hand panel), No-photo (a) (middle), and No-photo (b) (right-hand) images.** Plotted are averages for subbands and transform levels, denoted subband:level (e.g., the bar labeled LH:2 presents distribution of step-skip decisions applied when computing *LH* subband at level 2 of SS-DWT).

*LH* subbands) are nearly always skipped; the prediction step for the *HH* subband is also frequently skipped. For No-Photo (b) and Photo images, the prediction steps for the *H* and *HL* subbands are very frequently performed. The above described characteristics of No-photo (a) and other images is generally most pronounced for the lowest transform level that affects the highest number of samples and therefore has the greatest impact on the overall bitrate.

Based on the above observations of BH effects, we defined a revised heuristic (RH) by modifying the step B. We use the step A outcome to classify an image as No-photo (a) or other and then employ a suitable reduced complexity variant of step B, i.e., B1 for No-photo (a) or B2 for No-photo (b) and Photo:

A) Perform JPEG 2000 compression of an image using DWT and using SS-DWT with skipping all steps. Then, to all subbands at all transform levels, assign the decision that resulted in a better bitrate.

B1) If skipping was selected in step A, then for each transform level (starting from level 1) and for subbands at the specific level analyzed in the *H*, *HL*, and *HH* order, check if the overall bitrate is improved by inverting the decision for this subband and level; if it does, then invert the decision.

B2) If skipping was not selected in step A, then for each transform level (starting from level 1) and for subbands at the specific level analyzed in the *L*, *HH*, *LL*, and *LH* order, if the subband is different to *LH*, or performing the step is selected for the *HH* subband, then check if the overall bitrate is improved by inverting the decision for this subband and level; if it does, then invert the decision. If skipping was selected for *HH*, then select skipping for *LH*.

The RH computational time complexity is worse in the case of selecting to perform the lifting steps in step A of the heuristic. For the RH($n$) it is smaller than (using assumptions and symbols as in [Eq 8](#)):

$$\left(2 + \frac{10}{3}n\right)T_E + \left(1 + \frac{29}{12}n\right)T_D. \tag{9}$$

Two fixed SS-DWT variants were also defined. In SS-DWT FIX1, we skip all the update steps; in FIX2, we additionally skip prediction for the *HH* subband (and consequently the reorder for *HH* and *LH*). Compared to unmodified DWT, complexities of these variants are $T_D/2$ and $3T_D/8$, respectively. In a general case, SS-DWT is not compliant with the JPEG 2000 standard. However, variants FIX1 and FIX2 may be obtained by exploiting JPEG 2000 part 2 extensions defined in Annexes F and H of the standard [2, 5]. Note, that the JPEG 2000 standard defines as normative the syntax of the code stream containing the compressed image and the process of decoding it, whereas encoding procedures are included informatively only. The compression algorithm is compliant with the JPEG 2000 if it outputs a code stream that, by a JPEG 2000-compliant decompressor, is correctly decompressed.

The extension defined in Annex H allows specifying arbitrary wavelet kernels. The FIX1 variant may be obtained by simply defining a kernel, that uses regular prediction step of the reversible 5×3 DWT kernel ([Eq 1](#)) and skips the update by defining it to be $s_x \leftarrow s_x$. Since the arbitrary kernel is defined for a given tile (of a given component), the more complex variants of SS-DWT cannot be obtained this way.

The extension defined in Annex F of JPEG 2000 part 2 describes arbitrary decompositions of tile-components. Among others, at each decomposition level we may skip performing 1D-DWT in horizontal or vertical direction. As opposed to regular DWT that results in 4

**Table 3. Effects of SS-DWT on lossless JPEG 2000 bitrates for BH, RH, fixed SS-DWT variants, and for choosing among fixed transform variants.**

| Transform variant | Complexity | Time rel. | Images | | | | |
| --- | --- | --- | --- | --- | --- | --- | --- |
| | | | Photo | No-photo | No-photo (a) | No-photo (b) | All |
| $r_{DWT}$ | $T_D + T_E + T_R$ | 1.00 | 3.9975 | 2.9162 | 3.2882 | 2.7348 | 3.6395 |
| $\Delta r_{BH(1)}$ | $5.33\ T_D + 7.33\ T_E + T_R$ | 6.03 | -0.75% | -15.25% | -32.26% | -6.96% | -5.55% |
| $\Delta r_{BH(2)}$ | $9.67\ T_D + 12.67\ T_E + T_R$ | 10.30 | -0.86% | -15.66% | -32.28% | -7.54% | -5.76% |
| $\Delta r_{RH(1)}$ | $3.42\ T_D + 5.33\ T_E + T_R$ | 4.41 | -0.97% | -15.66% | -32.26% | -7.56% | -5.84% |
| $\Delta r_{RH(2)}$ | $5.83\ T_D + 8.67\ T_E + T_R$ | 7.07 | -0.98% | -15.73% | -32.28% | -7.65% | -5.86% |
| $\Delta r_{NO\_DWT}$ | $0.00\ T_D + 1.00\ T_E + T_R$ | 0.95 | 25.59% | 18.77% | -22.65% | 38.98% | 23.33% |
| $\Delta r_{DWT,\ NO\_DWT}$ | $1.00\ T_D + 2.00\ T_E + T_R$ | 1.76 | -0.01% | -7.43% | -22.65% | 0.00% | -2.47% |
| $\Delta r_{FIX1}$ | $0.50\ T_D + 1.00\ T_E + T_R$ | 0.97 | 0.15% | -10.94% | -22.07% | -5.51% | -3.52% |
| $\Delta r_{FIX1,\ DWT}$ | $1.22\ T_D + 1.75\ T_E + T_R$ | 1.58 | -0.46% | -11.14% | -22.07% | -5.80% | -3.99% |
| $\Delta r_{FIX1,\ DWT,\ NO\_DWT}$ | $1.22\ T_D + 2.75\ T_E + T_R$ | 2.34 | -0.47% | -13.10% | -28.05% | -5.80% | -4.65% |
| $\Delta r_{FIX2}$ | $0.38\ T_D + 1.00\ T_E + T_R$ | 0.97 | 0.30% | -14.11% | -29.27% | -6.72% | -4.47% |
| $\Delta r_{FIX2,\ DWT}$ | $1.19\ T_D + 2.00\ T_E + T_R$ | 1.77 | -0.39% | -14.25% | -29.27% | -6.93% | -4.98% |
| $\Delta r_{FIX2,\ DWT,\ NO\_DWT}$ | $1.19\ T_D + 3.00\ T_E + T_R$ | 2.53 | -0.40% | -15.13% | -31.93% | -6.93% | -5.27% |
| $\Delta r_{FIX1,\ FIX2}$ | $0.50\ T_D + 1.67\ T_E + T_R$ | 1.48 | -0.24% | -14.53% | -29.28% | -7.33% | -4.97% |
| $\Delta r_{FIX1,\ FIX2,\ DWT}$ | $1.22\ T_D + 2.42\ T_E + T_R$ | 2.09 | -0.62% | -14.61% | -29.28% | -7.45% | -5.25% |
| $\Delta r_{FIX1,\ FIX2,\ DWT,\ NO\text{-}DWT}$ | $1.22\ T_D + 3.42\ T_E + T_R$ | 2.84 | -0.62% | -15.48% | -31.93% | -7.45% | -5.54% |

$r_{DWT}$–average lossless JPEG 2000 bitrate for 3-level unmodified DWT (bpp); $\Delta r_{variant\_list}$–average bitrate change obtained using the best out of the listed variants; Complexity–compression process computational time complexity including complexity of heuristic or variant selection; Time rel.–compression time relative to time of unmodified JPEG 2000 compression.

subbands, performing 1D-DWT in just 1 direction results in 2 subbands. The JPEG 2000 part 2 entropy coding of such subbands may be more efficient than that of part 1 applied to SS-DWT subbands because the coder is aware of the actual decomposition applied. The transform at level $t+1$ is applied to the low-pass subband obtained with a $t$-level transform and we may specify decomposition structure individually for each level. At each decomposition level of FIX2 we need to skip 1D-DWT for $LH$ and $HH$ subbands and perform it for $LL$ and $HL$. This is not directly supported by Annex H, but the 1-level FIX2 may be substituted by a 2-level part 2 compliant decomposition that at level 1 is performed only in vertical and at level 2 only in horizontal direction. Therefore, the 3-level FIX2 SS-DWT variant may be obtained by using an arbitrary DWT kernel with the skipped update step and a 6-level arbitrary decomposition that at odd levels is performed only in vertical direction and at even levels only in horizontal direction.

In Table 3, we present average improvements of JPEG 2000 bitrate due to BH, RH, FIX1, and FIX2 variants of SS-DWT as well as results for combinations of all or some of variants: FIX1, FIX2, unmodified DWT, and compression with the DWT stage skipped (NO-DWT), obtained by selecting the best variant for each image. In each case we also report the computational time complexity (Complexity), which is expressed using the complexity of the elements of the unaltered JPEG 2000 process ($T_D$–DWT computation, $T_E$–encoding of transformed image, $T_R$–remaining JPEG 2000 operations [e.g., file i/o]). For combinations of variants, we take into account identities that allow reducing complexity of determining the better one. For example, subbands $H$ and $HH$ at level 1 of the DWT and SS-DWT FIX1 variant are identical and may be computed (and encoded in the case of the $HH$ subband) only once. The reported compression time (Time rel.) is expressed as relative to the time of an unmodified JPEG 2000; it is calculated based on the above complexity and the actual execution times of elements of JPEG 2000, which were measured on a computer system used in this research and averaged for several large images from the test-set (Table 4).

Table 4. Execution times of unmodified JPEG 2000 elements and of entropy calculation.

| Description | Time (ms per $10^6$ symbols) | Percentage of $T_J$ |
|---|---|---|
| Unmodified JPEG 2000 compression ($T_J$) | 543.7 | 100% |
| DWT transform | 29.0 | 5.34% |
| Entropy coding | 411.9 | 75.77% |
| Remaining operations | 102.7 | 18.89% |
| Entropy calculation | 4.2 | 0.78% |

doi:10.1371/journal.pone.0168704.t004

We can see for RH, that only 1 iteration of step B is sufficient, as the 2nd iteration gives a negligible further improvement of 0.02 percentage points on average for the whole set. Looking at RH(1) improvements for individual images (Fig 3), we see that greater improvements are more likely for better-compressible images (both Photo and No-photo) and that below certain lossless JPEG 2000 bitrate (about 3 bpp), SS-DWT improves the bitrate of each image. Although compression with RH is faster than with BH by roughly one third for the same number of iterations, it results in better overall improvements achieved in a smaller number of iterations. BH probably tends to find worse local optimums, which may be due to testing one subband at a time, while subbands are interdependent. The RH average bitrate improvement of 5.84% for the whole set and of nearly 1% for Photo images is obtained at the cost of increased complexity of the compression process. On our computer system, compression with RH(1) requires 4.4 times more time than unmodified JPEG 2000. Such a cost may be too high for certain practical applications; we reduce it by using fixed variants discussed below and by using estimation of entropy coding effects, as described in the following subsection. Note that as compared to DWT, the decompression time decreases a little due to reducing the number of lifting steps needed to compute SS-DWT.

Fixed FIX1 and FIX2 variants of SS-DWT improve the compression process complexity and the average bitrate. The bitrate is improved by roughly 3.5% to 4.5%, respectively, which is an effect of improved bitrates of No-photo images and of little worsened bitrates of Photo images. The latter are worsened by 0.15% and 0.30%, respectively, which is an averaged effect of much larger bitrate changes (both positive and negative) of individual images; see Fig 3B for FIX1 effects on individual images. FIX1 results also indicate that in the average case, the update step is either of low importance or harmful for DWT-based lossless JPEG 2000 compression.

To obtain bitrate improvements for both Photo and No-Photo images, we must evaluate the effects of multiple variants. When choosing between using unmodified DWT and NO_DWT, at the increased cost of double encoding of the image (i.e., of increasing the overall compression time by 76%), we get an average bitrate improvement of roughly 2.5%. The improvement is obtained without using SS-DWT; therefore, the compressed image is compliant with the JPEG 2000 baseline standard [1]. The above result may be improved both in terms of complexity and bitrate. Choosing between FIX1 and FIX2 results in an average improvement of roughly 5% at the smaller increase in cost. Extending the choice with DWT or DWT and NO_DWT, we get bitrate improvements closer to those obtained using RH and (to a smaller extent) compression times closer to RH. Image compressed this way complies with JPEG 2000 part 2 [5].

## Entropy-based estimation of subband encoding effects

Entropy coding is relatively slow. For the unmodified JPEG 2000 implementation we used, about 75% of the compression process time was spent on entropy coding. Therefore, we tested memoryless entropy of transformed subbands, $H_0$, as an estimator of entropy coding effects. In Table 5, we present average bitrate improvements obtained by using entropy estimation for

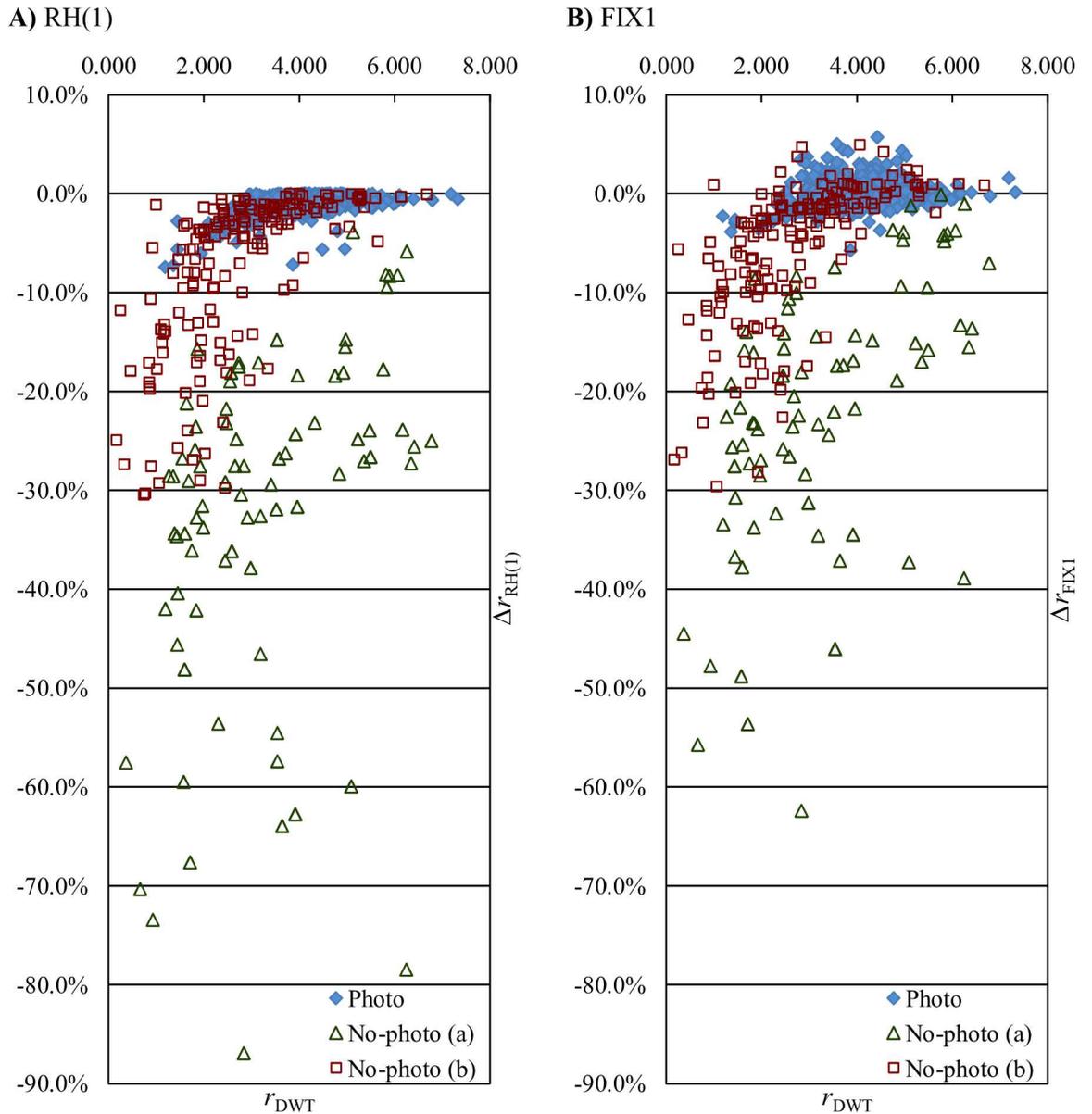

**Fig 3. SS-DWT bitrate changes for individual images.** $r_{DWT}$—lossless JPEG 2000 bitrate for 3-level unmodified DWT (bpp); $\Delta r_{variant}$—SS-DWT bitrate change obtained with use of SS-DWT variants: RH(1) (A) and FIX1 (B).

doi:10.1371/journal.pone.0168704.g003

selecting of steps to be skipped by the heuristic and for selecting among fixed transform variants. Obviously, then the actual (not estimated) compression was performed using the already determined fixed variant or step-skip decisions. Note, that if an estimation is used to choose the fixed DWT variant or to select decisions by the heuristic, then after finding the best variant or decision set we do not have the transformed samples already encoded, and during the actual compression we need to perform the actual entropy coding. Complexities in Table 5 are expressed using the complexity of the elements of the unaltered JPEG 2000 process ($T_D$—DWT computation, $T_E$—encoding of transformed image, $T_R$—remaining JPEG 2000 operations [e.g., file i/o]) and by $T_{H0}$ we denote the complexity of the bitrate estimation for the whole transformed image. We also report the compression time relative to time of unmodified JPEG 2000 $r_{DWT}$—average lossless JPEG 2000 bitrate for 3-level unmodified DWT (bpp); $\Delta r_{H0\ variant\_list}$—average bitrate change obtained for listed variants by employing entropy estimation for selecting fixed variants or steps to be skipped; Complexity—compression process computational time complexity including complexity of the heuristic or the variant selection; Time rel.—compression time relative to time of unmodified JPEG 2000.

**Table 5. Effects of SS-DWT on lossless JPEG 2000 bitrates for entropy estimation-based BH, RH, and choosing among fixed transform variants.**

| Transform variant | Complexity | Time | Images | | | | |
|---|---|---|---|---|---|---|---|
| | | rel. | Photo | No-photo | No-photo(a) | No-photo(b) | All |
| $r_{DWT}$ | $T_D + T_E + T_R$ | 1.00 | 3.9975 | 2.9162 | 3.2882 | 2.7348 | 3.6395 |
| $\Delta r_{H0}$ BH(1) | $5.33\,T_D + 7.33\,T_{H0} + T_E + T_R$ | 1.29 | -0.72% | -13.46% | -27.45% | -6.64% | -4.94% |
| $\Delta r_{H0}$ BH(2) | $9.67\,T_D + 12.67\,T_{H0} + T_E + T_R$ | 1.56 | -0.75% | -13.50% | -27.79% | -6.53% | -4.97% |
| $\Delta r_{H0}$ RH(1) | $3.42\,T_D + 5.33\,T_{H0} + T_E + T_R$ | 1.17 | -0.79% | -13.84% | -27.59% | -7.13% | -5.11% |
| $\Delta r_{H0}$ RH(2) | $5.83\,T_D + 8.67\,T_{H0} + T_E + T_R$ | 1.33 | -0.79% | -13.65% | -27.62% | -6.83% | -5.05% |
| $\Delta r_{H0}$ NO_DWT, DWT | $1.00\,T_D + 2.00\,T_{H0} + T_E + T_R$ | 1.02 | 0.00% | -5.43% | -17.35% | 0.38% | -1.80% |
| $\Delta r_{H0}$ FIX1, DWT | $1.22\,T_D + 1.75\,T_{H0} + T_E + T_R$ | 1.03 | -0.44% | -11.13% | -22.07% | -5.79% | -3.98% |
| $\Delta r_{H0}$ FIX1, DWT, NO_DWT | $1.22\,T_D + 2.75\,T_{H0} + T_E + T_R$ | 1.03 | -0.44% | -11.54% | -23.54% | -5.68% | -4.12% |
| $\Delta r_{H0}$ FIX2, DWT | $1.19\,T_D + 2.00\,T_{H0} + T_E + T_R$ | 1.03 | -0.12% | -13.76% | -29.27% | -6.19% | -4.63% |
| $\Delta r_{H0}$ FIX2, DWT, NO_DWT | $1.19\,T_D + 3.00\,T_{H0} + T_E + T_R$ | 1.03 | -0.12% | -13.39% | -28.46% | -6.03% | -4.51% |
| $\Delta r_{H0}$ FIX1, FIX2 | $0.50\,T_D + 1.67\,T_{H0} + T_E + T_R$ | 0.99 | 0.03% | -13.99% | -29.04% | -6.64% | -4.61% |
| $\Delta r_{H0}$ FIX1, FIX2, DWT | $1.22\,T_D + 2.42\,T_{H0} + T_E + T_R$ | 1.03 | -0.47% | -14.10% | -29.04% | -6.81% | -4.98% |
| $\Delta r_{H0}$ FIX1, FIX2, DWT, NO-DWT | $1.22\,T_D + 3.42\,T_{H0} + T_E + T_R$ | 1.04 | -0.47% | -13.77% | -28.34% | -6.66% | -4.87% |

(Time rel.). In Fig 4, the effects on average bitrate change obtained with use of various fixed, entropy estimation, and bitrate-based SS-DWT variants are plotted against the above time.

Generally, estimation-based SS-DWT bitrate improvements are smaller than actual bitrate-based ones, but they are alike and they are obtained at a greatly reduced cost. For Photo images (Fig 4B), bitrate improvements due to RH and BH (both entropy estimation and bitrate-based) result in better compression ratios than single or combined fixed variants. For No-photo images (Fig 4C), the improvements are several times greater; the best combinations of fixed variants obtain average bitrate improvements that are close to improvements obtained by the heuristic (both for estimation-based and independently for actual bitrate-based decisions); however, all the entropy estimation-based variants obtain worse average bitrates compared to the fixed FIX2 variant.

Looking at entropy estimation effects for individual images in the case of RH(1) (Fig 5A), and when employing entropy estimation for choosing between FIX1, FIX2, and DWT (Fig 5B), we notice that below certain bitrate of unmodified JPEG 2000, bitrates of almost all images are improved by the above entropy estimation-based variants. I.e., these variants perform similarly to the actual bitrate-based RH(1) (Fig 3A). With respect to average bitrate improvements and standard deviations of bitrate improvements (see Table 6) these variants are similar in the case of non-photographic images, whereas for photographic images choosing between FIX1, FIX2, and DWT is clearly worse. However, as the distribution of bitrate improvements is skewed, the standard deviation is not a good basis for determining how often (and whether) a certain variant of SS-DWT may result in bitrate worsening. Except for the small cost of encoding the decisions on which steps to skip, the actual bitrate-based RH(1) does not worsen bitrate of any image. This heuristic allows a change in SS-DWT only if such a change improves the actual JPEG 2000 bitrate. Estimation-based variants may result in worsening of bitrates of some images, but such cases are not frequent. The estimation-based RH(1) worsened bitrates of 7.4% of Photo images and 1.8% of No-photo (b) images, whereas the estimation-based choosing between FIX1, FIX2, and DWT expanded 5.4% of Photo images and 3.0% of No-photo (b) images.

The estimation effects (see Table 5) are satisfactory from a practical standpoint. The greatest average bitrate improvement of 5.11% was obtained for the RH(1) at the cost of increasing the compression process time by 17%. For Photo images, most of the RH(1) improvement obtained using the actual entropy coding was also obtained by the use of entropy estimation (0.97% vs. 0.79%). The level of improvement obtained by SS-DWT for continuous-tone photographic images is not large; however, it is quite good as compared to recently reported improvements obtained using much more complex modifications of JPEG 2000 and DWT [10]. In the above-mentioned study for medical CT, MRI, and US data, which were treated as 3D volumes and as collections of 2D images, the bitrates were improved on average by below 1% by applying block-based intra-band prediction or direction-adaptive DWT.

Estimation may lead to decreased bitrate improvements in cases where the potential of improvement still exists, and by making decisions based on the actual bitrate we obtain greater improvements. Performing a $2^{nd}$ iteration of RH (compare rows labeled $\Delta r_{H0\ RH(1)}$ and $\Delta r_{H0\ RH(2)}$ in Table 5) or extending with NO_DWT the selection between FIX2 and DWT ($\Delta r_{H0\ FIX2,\ DWT}$ and $\Delta r_{H0\ FIX2,\ NO\_DWT}$) or between FIX1, FIX2, and DWT

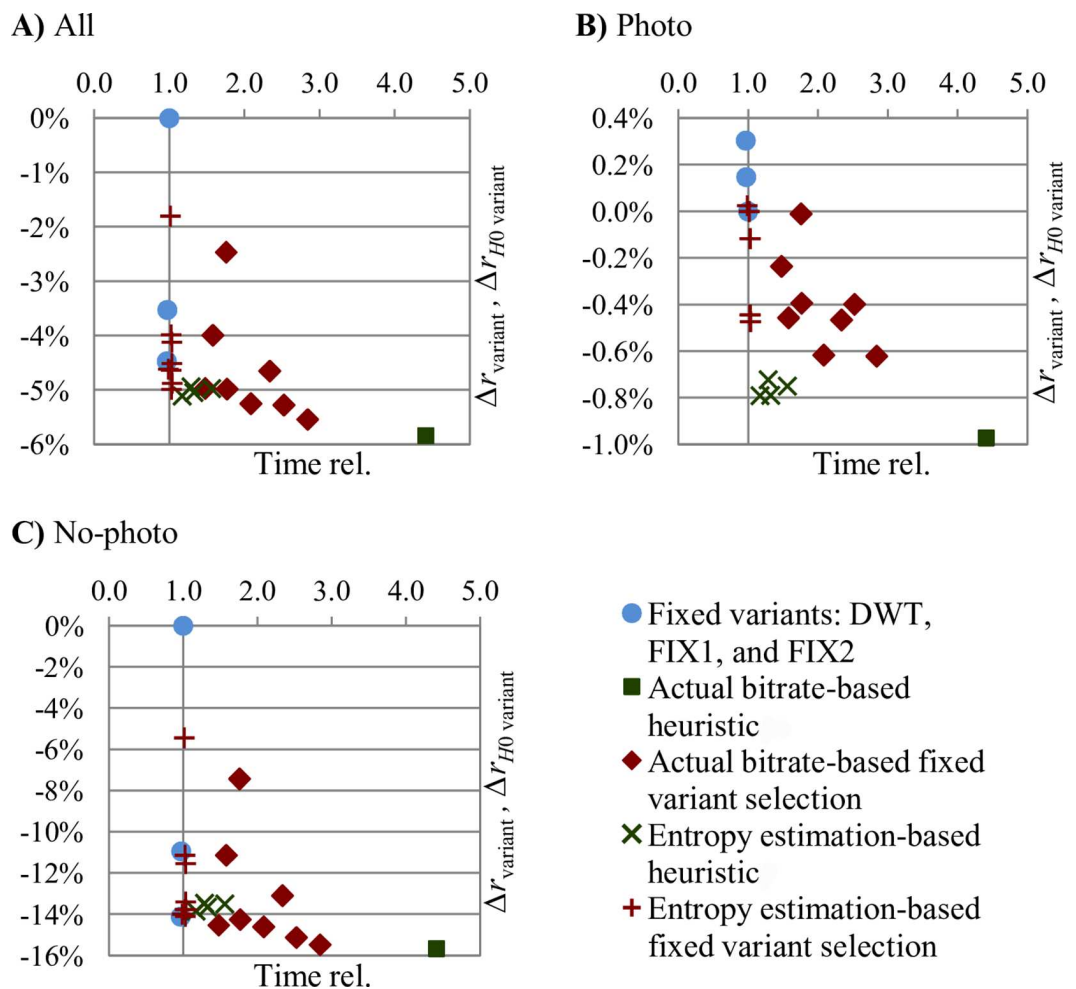

**Fig 4. Entropy estimation-based and actual JPEG 2000 bitrate-based SS-DWT average bitrate changes ($\Delta r$)** plotted against compression time relative to unmodified JPEG 2000 (Time rel.). NO_DWT and variants over 5 times slower than JPEG 2000 are not plotted.

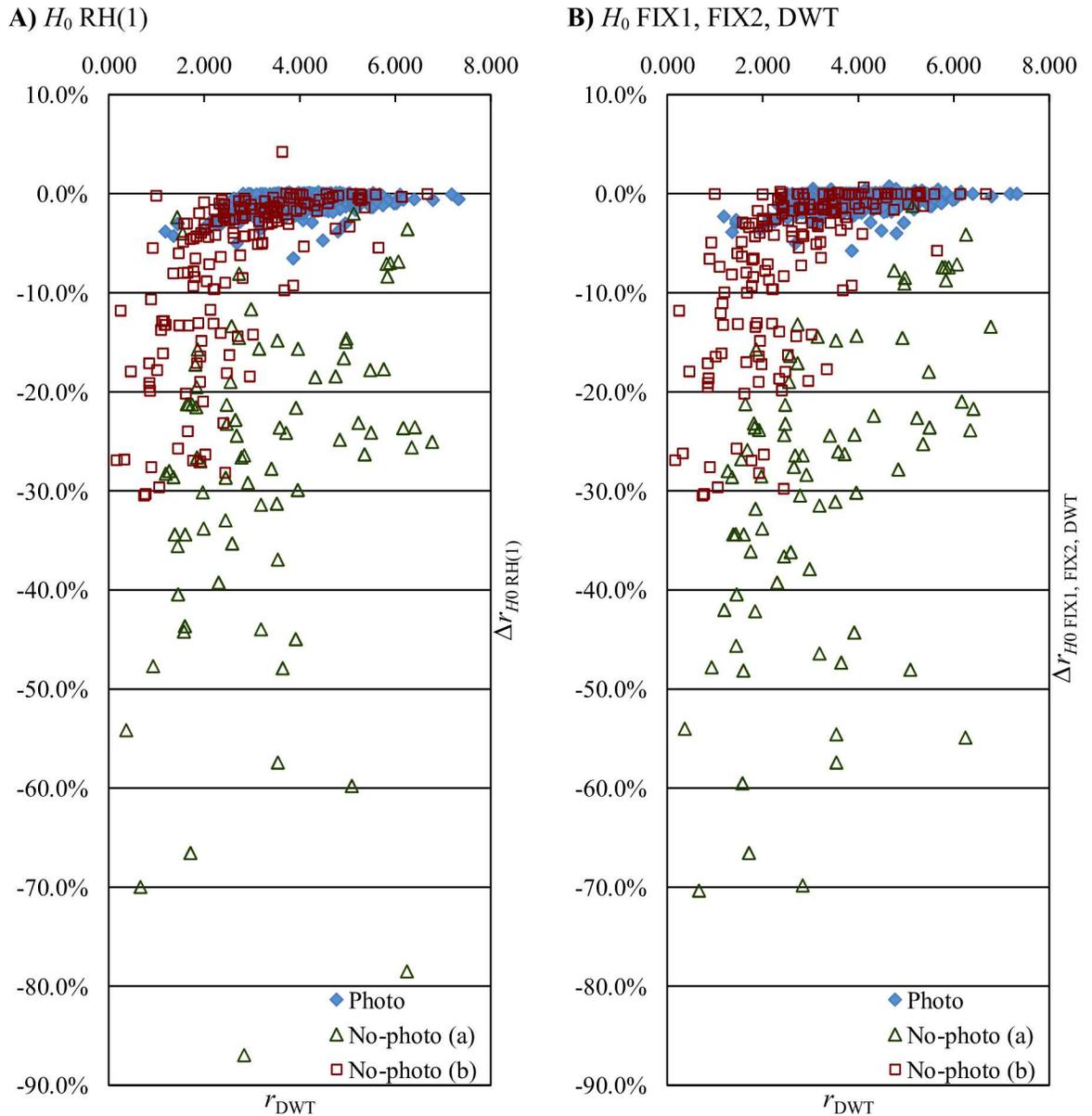

**Fig 5. Entropy estimation-based SS-DWT bitrate changes for individual images.** $r_{DWT}$−lossless JPEG 2000 bitrate for 3-level unmodified DWT (bpp); $\Delta r_{H0\ variant}$−SS-DWT bitrate change obtained with the use of the entropy estimation-based variants: RH(1) (A) and choosing between FIX1, FIX2, and DWT (B).

doi:10.1371/journal.pone.0168704.g005

($\Delta r_{H0\ FIX1,\ FIX2,\ DWT}$ and $\Delta r_{H0\ FIX1,\ FIX2,\ DWT,\ NO-DWT}$) results in worse entropy estimation-based improvements of No-photo images. These observations, along with the fact that estimation effects are generally worse to using actual bitrates, suggest that the estimation might be improved.

In other studies [4, 17, 18], the memoryless entropy of the color image component prediction error obtained using the nonlinear MED predictor [24] was used for selecting a color space transform; it was found to be a very efficient estimator of coding effects of JPEG 2000 as well as of other lossless image compression algorithms that exploit context-adaptive entropy coding: JPEG-LS and JPEG XR [25, 26]. However, that estimation was applied to data that had

**Table 6. Variability of bitrate improvements of Photo, No-photo (a) and No-photo (b) images for SS-DWT variants presented in Figs 3 and 5.**

| Transform variant | Images | | |
|---|---|---|---|
| | **Photo** | **No-photo (a)** | **No-photo (b)** |
| $\Delta r_{RH(1)}$ | -0.97%(1.06%) | -32.26%(17.39%) | -7.56%(8.10%) |
| $\Delta r_{FIX1}$ | 0.15%(1.40%) | -22.07%(13.71%) | -5.51%(7.14%) |
| $\Delta r_{H0\ RH(1)}$ | -0.79%(0.87%) | -27.59%(16.51%) | -7.13%(8.12%) |
| $\Delta r_{H0\ FIX1,\ FIX2,\ DWT}$ | -0.47%(0.80%) | -29.04%(15.51%) | -6.81%(8.12%) |

$\Delta r_{variant\_list}$, $\Delta r_{H0\ variant\_list}$—average bitrate change reported with standard deviation (in parenthesis) obtained for the best out of the listed actual bitrate-based and entropy estimation-based variants.

doi:10.1371/journal.pone.0168704.t006

not been already transformed into subbands of different characteristics. We suppose, that better and still fast estimates of JPEG 2000 subband-dependant context entropy coding of SS-DWT subbands may be obtained by using conditional entropy or subband-dependant predictors, which is an interesting area for future research. As the characteristics of SS-DWT subbands are in some cases closer to the characteristics of untransformed images than to DWT-transformed subbands, we also suppose that the intra-band prediction effects may be better in the case of SS-DWT. This way, a compression scheme could be obtained, that lies in-between typical transform-based algorithms, like JPEG 2000, and predictive ones, like JPEG-LS.

Recall, that the compression algorithm is compliant with the JPEG 2000 standard if it outputs a code stream that, by a JPEG 2000-compliant decoder, is correctly decompressed. All the fixed variants (FIX1, FIX2, DWT, NO-DWT) may be encoded in compliance with JPEG 2000 part 2 and then decoded with any decoder compliant with this standard. Thus the entropy-estimation based selection of fixed variants is compliant with the standard. Obtaining the variant selection in practice requires either to extend the regular coder with the entropy estimation-based selection routines, or to perform the selection beforehand using a separate tool. The cost of the latter case may be little higher, compared to the cost reported in Table 5, due to additional costs of file input and performing SS-DWT both by the separate tool and the coder. Note also, that using just the unmodified coder, the actual bitrate-based fixed variants reported in Table 3 may be simply obtained by multiple runs of compression using different variants and then by picking the best one; in such a case the cost will be higher than reported in Table 3.

The selecting between FIX1, FIX2, and unmodified DWT based on entropy estimation (see Fig 5B and row labeled $\Delta r_{H0\ FIX1,\ FIX2,\ DWT}$ in Table 5) appears the most interesting variant from a practical standpoint. This variant is compliant with JPEG 2000 part 2 and it results in an average bitrate improvement of roughly 5% for the entire test-set, whereas the overall compression time gets only 3% greater than that of unmodified JPEG 2000. Bitrates of Photo and No-photo images were improved by roughly 0.5% and 14%, respectively. A little better improvements, overall and especially for Photo images, were attained using the estimation-based RH(1) that also is very fast, but is not compliant with the JPEG 2000 standard. Larger bitrate improvements may be attained at a significantly increased cost of using the actual bitrate instead of estimated one and by applying RDLS.

## Conclusions and future work

In order to improve bitrates of lossless JPEG 2000, we proposed the SS-DWT, which was obtained from DWT by skipping selected steps of its computation. A heuristic, image-adaptive determination of steps to be skipped was initially done by BH and involved the relatively slow

JPEG 2000 entropy coding. The SS-DWT bitrate improvements were better than we expected, as they exceeded the improvements obtained previously by using more complex RDLS-DWT. We found that even greater improvements might be attained by combining both methods, however, at a greater cost.

Then from a practical standpoint, based on BH outcomes, we proposed RH with a lower computational time complexity and defined fixed SS-DWT variants that were compliant with the JPEG 2000 part 2 standard. To further reduce the cost of bitrate improvement, we used subband entropy as an estimator of JPEG 2000 encoding effects for the heuristic and for choosing among fixed variants. We found that SS-DWT allows for significant bitrate improvements of non-photographic images. At a cost of compression time 4.4 times greater than that of unmodified JPEG 2000, by using the SS-DWT step-skip decisions obtained with RH(1) based on actual JPEG 2000 bitrates, we attain an average bitrate improvement of nearly 6%. Bitrates of photographic and non-photographic images are improved by almost 1% and 16% respectively. This variant does not worsen the bitrate of any image and improves bitrates of all images that by unmodified JPEG 2000 are compressed to below about 3 bpp. Using the entropy estimation of coding effects, the overall time of compression with RH(1) gets effectively lowered to just 17% greater than the time of unmodified JPEG 2000, whereas the bitrate is lowered by over 5% on average. The most interesting results from a practical standpoint, however, are obtained by employing the entropy estimation for selecting among fixed SS-DWT variants. This way we get the JPEG 2000 part 2 compliant compression scheme that, for our research implementation and test-set, resulted in an average bitrate improvement of almost 5% for the entire test-set and over 14% for non-photographic images, whereas the overall compression time was only 3% greater than that of unmodified JPEG 2000.

In a general case, the characteristics of images transformed by SS-DWT are different to what is expected by the baseline JPEG 2000 (part 1) subband-dependant context entropy coder, and therefore the transformed image is not encoded in the most efficient way. Fixed variants of SS-DWT are compliant with the JPEG 2000 part 2 standard, but the implementation we used in experiments was compliant with part 1 only. Despite that, for most images we obtained the greatest bitrate improvements by skipping of some steps. This presents a possibility to further improve the bitrate by employing an implementation of part 2 of JPEG 2000, which could be modified to make the coder better match the general-case SS-DWT subband decomposition and characteristics of the subbands. The effectiveness of the entropy estimation of coding effects that we employed may be probably improved by using conditional entropy or subband-dependant predictors, which is a potential area for future research. In the ongoing research we investigate the intra-band prediction of SS-DWT subbands; the results obtained so far are encouraging. We also apply step skipping to the 3D-DWT employed by JPEG 2000 part 10 in lossless compression of volumetric medical data. We have already applied RDLS with step skipping to several lifting-based reversible color space transforms [13]. Finally, we plan to investigate the image adaptive selection of the color space transform modified using RDLS with step skipping and to combine such a transform with the JPEG 2000 compression exploiting step skipping, RDLS, and the intra-band prediction.

## Author Contributions

**Conceptualization:** RS.

**Formal analysis:** RS.

**Funding acquisition:** RS.

**Investigation:** RS.

**Methodology:** RS.

**Project administration:** RS.

**Software:** RS.

**Validation:** RS.

**Visualization:** RS.

**Writing – original draft:** RS.

**Writing – review & editing:** RS.